\documentclass[twocolumn,           
               showpacs,            
               preprintnumbers,     
               aps,                 
               prd,                 
               a4paper,             
               superscriptaddress,  
               unsortedaddress,     
               footinbib,           
               tightenlines,        
               floats               
               ]{revtex4}

\usepackage{graphicx}
\usepackage{latexsym}
\usepackage{amsmath,amssymb}        
\usepackage[draft=false]{hyperref}

\newcommand{\de}{\mathrm{d}}
\renewcommand{\(}{\left(}
\renewcommand{\)}{\right)}
\renewcommand{\[}{\left[}
\renewcommand{\]}{\right]}
\newcommand{\period}{\,\mathrm{.}}
\newcommand{\comma}{\,\mathrm{,}}
\newcommand{\reffig}[1]{Fig.~\ref{#1}}
\newcommand{\refeq}[1]{Eq.~(\ref{#1})}
\newcommand{\refeqs}[2]{Eqs.~(\ref{#1}) and (\ref{#2})}

\newcommand{\mpl}{m_\mathrm{Pl}}

\newcommand{\abs}[1]{\left\vert#1\right\vert}

\newcommand{\e}{\varepsilon}
\newcommand{\Dphi}{\Delta\phi}
\newcommand{\Ninf}{N_\mathrm{inf}}
\newcommand{\Nmax}{N_\mathrm{max}}
\newcommand{\D}{\mathcal{D}}
\newcommand{\as}{\alpha_\mathrm{S}}

\begin{document}


\title{From the production of primordial perturbations to the end of inflation}
\author{Micha\"el Malquarti}
\affiliation{Astronomy Centre, University of Sussex, 
             Brighton BN1 9QH, United Kingdom}
\author{Samuel M. Leach}
\affiliation{D\'epartement de Physique Th\'eorique, Universit\'e de Gen\`eve,
             24 quai Ernest-Ansermet, 1211 Gen\`eve 4, Switzerland}
\author{Andrew R. Liddle}
\affiliation{Astronomy Centre, University of Sussex, 
             Brighton BN1 9QJ, United Kingdom}
\date{\today} 
\pacs{98.80.Cq \hfill astro-ph/0310498}
\preprint{astro-ph/0310498}


\begin{abstract}
In addition to generating the appropriate perturbation power spectrum, an 
inflationary scenario must take into account the need for inflation to end 
subsequently. In the context of single-field inflation models where inflation 
ends by breaking of the slow-roll condition, we constrain the first and second 
derivatives of the inflaton potential using this additional requirement. We 
compare this with current observational constraints from the primordial spectrum 
and discuss several issues relating to our results.
\end{abstract}

\maketitle


\section{Introduction}

With the increasing precision of cosmological observations, inflation has become 
the favored candidate for explaining the origin of perturbations in the Universe 
\cite{infrev}. While some plausible scenarios have recently been introduced 
whereby adiabatic perturbations are generated after inflation from isocurvature 
perturbations laid down during inflation \cite{isoad}, the generation of 
adiabatic density perturbations during inflation remains the simplest one. 
However, although inflationary models give an excellent fit to most recent data 
including that of the Wilkinson Microwave Anisotropy Probe 
(\textsc{wmap})~\cite{wmap,Petal}, the perturbations are observable only over a 
fairly narrow range of scales, corresponding to about four orders of magnitude 
in wavenumber, thus allowing us to constrain only a small segment of the 
inflationary potential. Nevertheless, there is one further piece of information 
that can be brought into play \cite{infend}, which is that we know that 
inflation must come to an end soon after the observable perturbations are 
generated.

The literature describes three ways in which inflation might end. In the 
simplest scenario, requiring just a single scalar field, the logarithm of the 
potential driving inflation becomes too steep to sustain inflation, leading to 
the end of the slow-roll regime and usually giving way to a series of 
oscillations about a minimum in the potential. A second popular possibility is 
an instability, associated with a second scalar field, which removes the 
potential energy driving inflation; this is the key idea of the hybrid inflation 
paradigm~\cite{hyb}, where inflation ends by a phase transition. Much less 
discussed is a third possibility, that at some energy scale the underlying 
equations of motion are modified, an example being the steep inflation 
model~\cite{steep} where inflation is sustained only by corrections to the 
Friedman equation at high energies in a braneworld model, with inflation ending 
as the energy scale drops and these corrections become unimportant.

In the hybrid inflation case, the inflaton field is normally unaware of the 
existence of the instability until its onset, and the inflaton dynamics gives no 
clue as to when it might happen. In that case, we can expect no useful extra 
information from the need to end inflation. If the underlying equations can be 
modified, as in the steep inflation case, there are many ways in which this 
could happen and it is unlikely that any useful model-independent statements can 
be made. In this paper, we therefore restrict our attention to models with a 
single scalar field, in which inflation ends by breaking of the slow-roll 
condition. Our aim is to assess whether the requirement to end inflation imposes 
useful additional constraints on the inflaton potential, and to discover whether 
there are regions of parameter space permitted by the perturbation data which 
are ill-suited to a satisfactory end to inflation in this manner.

It was recently shown that there is a firm upper limit $\Nmax$ to the number of 
$e$-foldings $\Ninf$ before the end of inflation at which observable 
perturbations were generated~\cite{Ninf}. In this work, we aim to use 
the value of $\Nmax$ to set some conservative constraints on the first two 
derivatives of the inflaton potential in the context of single-field inflation, 
which can be compared to the region permitted by the observed perturbations. In 
other words we want to look at the generic predictions of single-field 
inflation by defining a region in the primordial power spectrum parameter space 
compatible with the paradigm. This goal is similar to that of analyses using the 
inflationary flow equations \cite{flow}, such as Peiris et al.~\cite{Petal} and 
Kinney et al.~\cite{KKMR} which are based on the method of Easther and 
Kinney~\cite{EK}, but as we will discuss our approach is different in its 
physical content and makes more restrictive assumptions about the shape of the 
potential.

The paper is organized as follows: Section~\ref{methodology} is devoted to 
stating our assumptions and explaining our methodology, Section~\ref{results} 
gives the results and analysis, and Section~\ref{discussion} is a general 
discussion on related issues. We assume $\mpl=1$ throughout.

\section{Methodology}
\label{methodology}

In this section we explain how we constrain the inflaton potential in the 
context of single-field inflation. The main idea is to test whether it is 
possible to build a reasonable model which takes into account the upper limit on 
the number of $e$-foldings $\Ninf$ before the end of inflation, given the shape 
of the potential in the range probed by observations.

Given an inflationary potential $V(\phi)$ and an initial value of the field 
$\phi_*$ (corresponding to the horizon crossing of a pivot scale $k_*$) we can 
compute a Taylor expansion of $V(\phi)$ around $\phi_*$. In the context of 
slow-roll and in the face of the current observational data, one does not expect 
more than the first two or three derivatives to play an important role in the 
range of scales probed by \textsc{cmb} observations and galaxy distribution 
surveys. Before going into more details, it is useful to note that the evolution 
of the field as a function of the number of $e$-foldings does not depend on the 
normalization of the potential. Therefore, throughout the paper we use the 
parameters $V'/V$, $V''/V$, $V'''/V$, etc. which are evaluated at $\phi_*$. 
Also, by convention we take the first derivative to be negative.

Now, let us introduce the set of slow-roll parameters~\cite{STEG,LLMS}
\begin{eqnarray}
\e_0     &=& \frac{H(0)}{H(N)} \comma \\
\e_{n+1} &=& \frac{\de \ln\abs{\e_n}}{\de N}\qquad\mathrm{for}\quad n\ge0
\comma
\end{eqnarray}
where $H$ is the Hubble parameter and $N$ is the number of $e$-foldings since 
the crossing of the horizon by the pivot scale $k_*$. This particular choice of 
definition is known as the horizon-flow parameters. We can compute (to 
first-order) $V'/V$ and $V''/V$ as functions of $\e_1$ and $\e_2$
\begin{eqnarray}
\frac{V'}{V} &\simeq&
-4\sqrt{\pi\e_1}
\comma \\
\frac{V''}{V} &\simeq&
4\pi(4\e_1-\e_2)
\comma
\end{eqnarray}
which relates the shape of the potential with the inflationary dynamics. 
Conversely, we can recover the slow-roll parameters from the derivatives of the 
potential
\begin{eqnarray}
\e_1 &\simeq& \frac{1}{16\pi}\(\frac{V'}{V}\)^2
\comma \\
\e_2 &\simeq& \frac{1}{4\pi}\[\(\frac{V'}{V}\)^2-\frac{V''}{V}\]
\period
\end{eqnarray}
These slow-roll parameters can then be related to primordial power spectrum 
parameters such as the scalar and tensor spectral indices 
$n_\mathrm{S}-1\simeq-2\e_1-\e_2$ and $n_\mathrm{T}\simeq-2\e_1$, the tensor to 
scalar ratio $R\simeq16\e_1$, etc. Therefore, we will interchangeably use any 
independent pair of these parameters. Note that since the constraint on the 
running $\as\simeq-2\e_1\e_2-\e_2\e_3$ is too weak at the moment (see 
Ref.~\cite{LL03} for comments on this issue), we will assume some theoretical 
prior on this parameter. It is important to stress that in this work we do not 
want to constrain higher derivatives; rather, we are trying to find a region of 
the $V'/V$--$V''/V$ space that is compatible with a large class of single-field 
inflation models.

Now, as explained above, our aim is to try to build a working single-field 
inflation model (i.e.~an inflaton potential) by expanding the potential as a 
Taylor series, fixing the first two derivatives to reasonable values, and then 
varying the higher-order derivatives using a random process. We set the 
following rules to define a working model:
\begin{enumerate}
\item The shape of the potential should be consistent with the constraints on 
the primordial perturbation power spectrum, as well as with the prior on the 
running $\as$.
\item The potential should either be convex ($V''/V>0$) throughout the evolution 
(large-field inflation) or at first concave ($V''/V<0$) and eventually convex 
(small-field inflation).
\item The number of $e$-foldings between the time the scale $k_*$ leaves the 
horizon and the end of inflation ($\e_1=1$) should be less than $\Nmax$.
\end{enumerate}

Concerning the first rule, we look at a region of the parameter space 
$V'/V$--$V''/V$ approximately consistent with observations of $n_\mathrm{S}-1$ 
and $R$ --- we will later contrast our results with actual constraints from 
\textsc{wmap} and 2d\textsc{f}. We initially impose the theoretical prior 
$-0.04<\as<0.02$. In anticipation that future observations will pin down the 
running more precisely, we then go on to examine successively the following 
subcases: (i) $-0.04<\as<-0.02$, (ii) $-0.02<\as<0$, and (iii) $0<\as<0.02$, so 
as to investigate how future constraints may affect the overall picture. As
\begin{equation}\label{v3}
\begin{split}
\frac{V'''}{V}
&\simeq \sqrt{\frac{\pi}{\e_1}}\(-24\e_1^2+18\e_1\e_2-3\e_2\e_3\) \\
&\simeq \sqrt{\frac{\pi}{\e_1}}\(-24(\e_1^2-\e_1\e_2)+3\as\) \comma
\end{split}
\end{equation}
these constraints on the running impose constraints on the third derivative of 
the potential. Then, we assume that higher derivatives are negligible in the 
range of values of the field corresponding to observed scales. Specifically, we 
impose
\begin{equation}
\label{const34}
\abs{\frac{1}{4!}\frac{V^{(4)}}{V}\Dphi^4}<
\abs{\frac{1}{3!}\frac{V'''}{V}\Dphi^3} \comma
\end{equation}
where $\Dphi$ is the distance run by the field when producing the observed 
perturbations (corresponding to roughly 7 $e$-foldings, 3.5 on each side of 
$\phi_*$). This means that the fourth-order term in the Taylor expansion is 
assumed not to overtake the third-order term until the field runs about twice 
the distance to the edge of the observed region. Note that as a result, the 
fourth derivative term cannot contribute significantly to the curvature of the 
potential inside the observed region. In addition, in order to find working 
models more easily, we use the practical recipe
\begin{equation}
\label{const45}
\abs{\frac{1}{5!}\frac{V^{(5)}}{V}\Dphi_{3,4}^5}<
\abs{\frac{1}{4!}\frac{V^{(4)}}{V}\Dphi_{3,4}^4} \comma
\end{equation}
where $\Dphi_{3,4}$ is the value of the field for which the fourth order term 
equals the third order term in the Taylor expansion. Note that if $V'/V$ and 
$V''/V$ are fixed, the uncertainty on $\as$ still allows $V'''/V$ to vary and 
therefore the possibility of having $V'''/V\simeq0$ (and hence 
\refeqs{const34}{const45} being too strong constraints) is avoided.

The second rule is assumed in order to maintain the simplicity of the model, as 
this is the main reason for considering single-field inflation. Also, most 
models in the literature are of this form. We consider Taylor expansions of 
third, fourth and fifth order. It is important to note that our class of models 
is broader than a set of polynomial potentials, since the expansion need only 
approximate the true potential over a limited range, with the order of the 
expansion reflecting the number of degrees of freedom we have to shape the 
potential in order to fulfil our set of rules. We checked that a fifth degree 
polynomial can fit a wide range of potentials from $\phi_*$ to the value of the 
field corresponding to the end of inflation.

We also investigated the effect of imposing the constraint that \mbox{$V=0$} at 
the minimum of the potential. However we found that this condition complicates 
the analysis without adding anything useful, since after inflation ends it is 
usually not hard for the potential to then shape itself to form a satisfactory 
minimum. In any case it is not our intention to address the post-inflationary 
dynamics.

Finally, concerning the third rule, inflation must end by breaking of the 
slow-roll condition ($\e_1=1$) and this should happen within a certain number of 
$e$-foldings $\Ninf<\Nmax$ after the scale $k_*$ crosses the horizon. The 
uncertainty on $\Ninf$ comes mainly from the reheating process, which can be 
very brief or alternatively can last until nucleosynthesis. Assuming 
instantaneous reheating and with $h\simeq0.72$, $\Omega_\mathrm{m}\simeq0.27$, 
the amplitude of scalar perturbations $A_\mathrm{S}\simeq2.3\times10^{-9}$ and 
the pivot scale $k_*\simeq0.01$Mpc$^{-1}$ (see Ref.~\cite{LL03}) we have
\begin{eqnarray}
\Nmax &\simeq& 60+\frac14\ln(\epsilon_1)-\frac14\ln(\D) \comma
\end{eqnarray}
where $\D$ is the drop in the energy density between the time the scale $k_*$ 
crosses the horizon and the time inflation ends. The discrepancy between this 
equation and the results given in Ref.~\cite{Ninf} comes from choosing a 
different scale as a starting point.

Our procedure is as follows: We fix the pair $V'/V$ and $V''/V$, randomly choose 
the higher derivatives, and then test the resulting potential against our set of 
rules. We repeat this step until we find a working model. If we cannot build a 
potential fulfilling the three rules stated above after a certain number of 
tries $n_\mathrm{tries}$, then we say that this pair of parameters is not 
consistent with single-field inflation, and move onto the next values of $V'/V$ 
and $V''/V$. We took $n_\mathrm{tries} = 300^{\nu}$, where $\nu$ is the number 
of degrees of freedom describing the potential after the first two derivatives 
have been chosen (i.e. $\nu=1$, $2$ or $3$), and tested other values to ensure 
that our results do not depend on this choice.

\section{Results}
\label{results}

\begin{figure}[t!]
\includegraphics[scale=0.39,angle=-90]{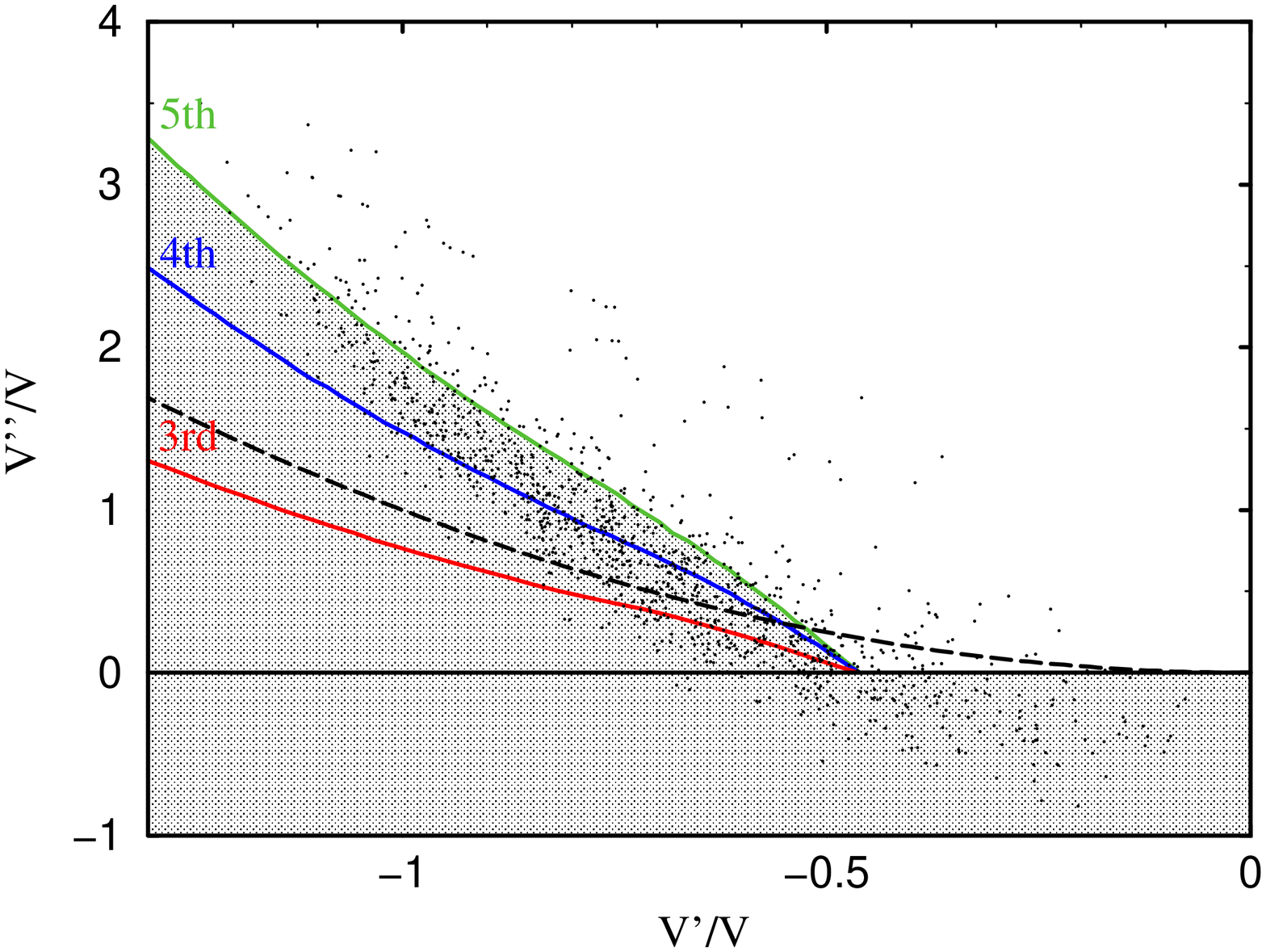}
\caption{Boundaries between the regions where it is possible to build a working 
model (shaded) and regions where single-field inflation is excluded. We assume 
$-0.04<\as<0.02$. The black line at $V''=0$ separates small-field and 
large-field models, and the dashed line shows $\e_2=0$. The other lines show the 
different orders in the expansion of the potential, with the labels always 
placed outside the allowed region. The dots show models from the Monte-Carlo 
Markov chain fitting the perturbation data.}
\label{plotmain}
\end{figure}

In this section we present our results as boundaries between allowed and 
excluded regions in the $V'/V$--$V''/V$ space and in the $(n_\mathrm{S}-1)$--$R$ 
space. In other words, we are seeking to make some falsifiable predictions for 
our class of single-field inflation models.

\begin{figure}[t!]
\includegraphics[scale=0.36,angle=-90]{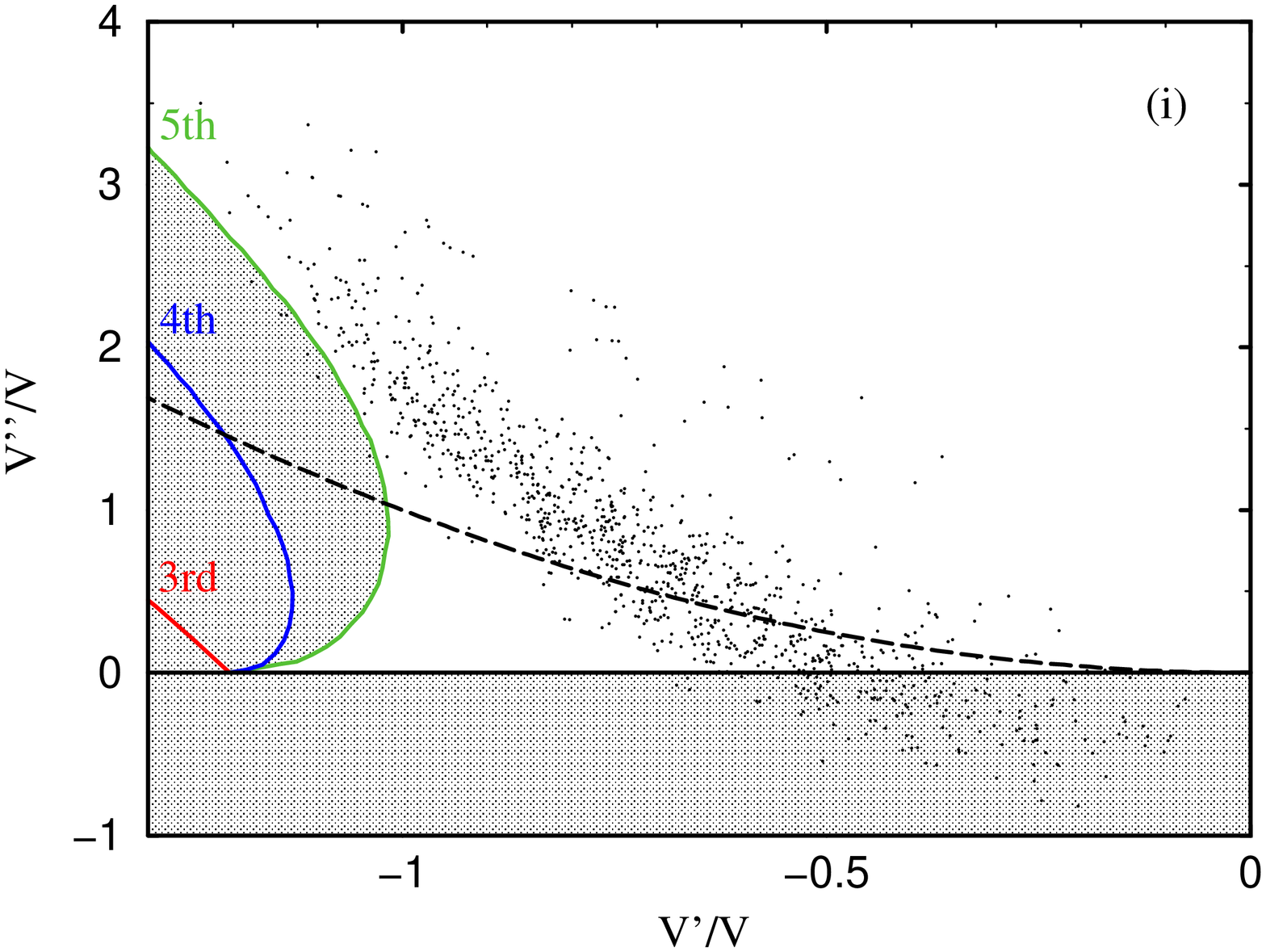}
\includegraphics[scale=0.36,angle=-90]{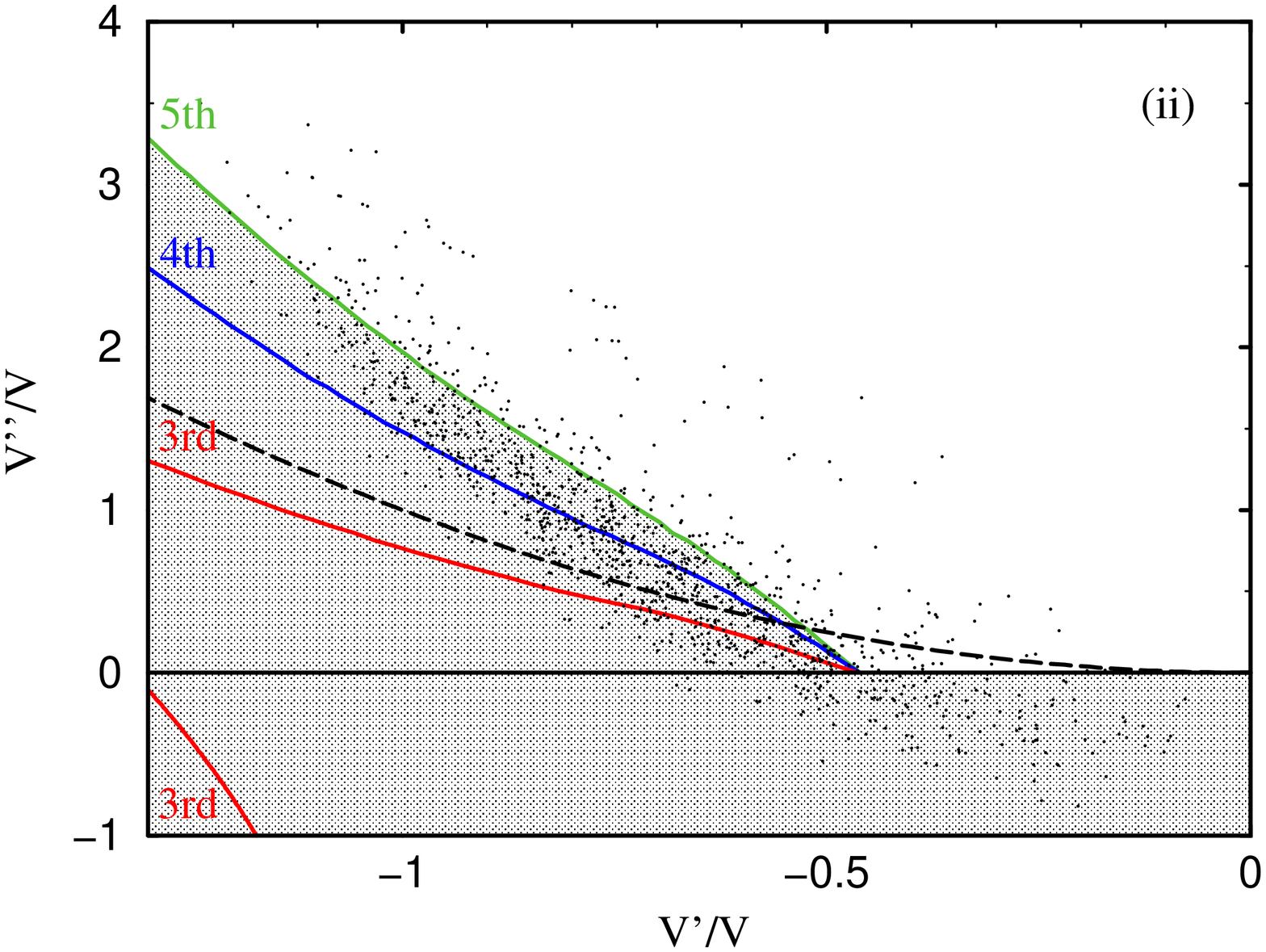}
\includegraphics[scale=0.36,angle=-90]{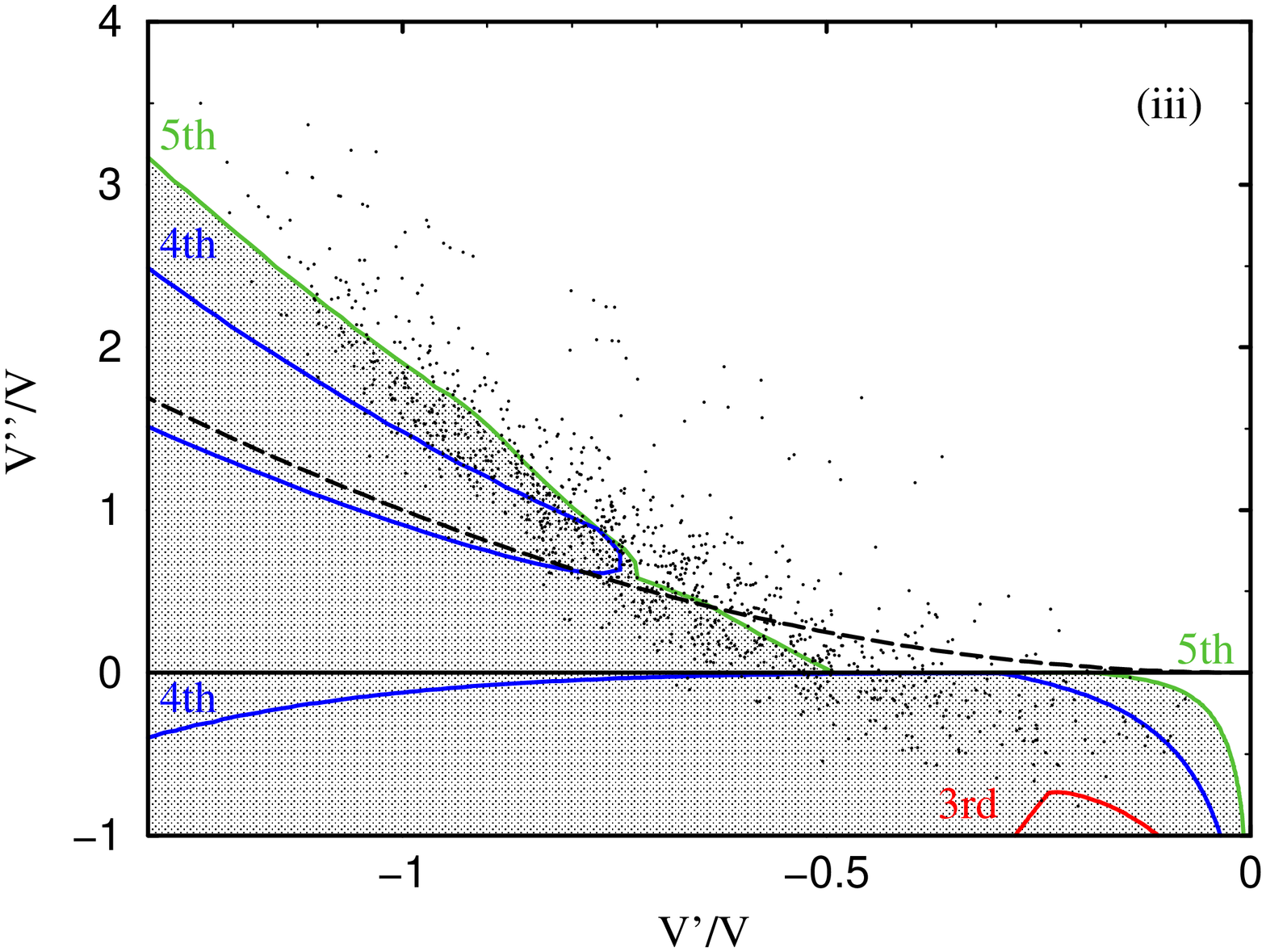}
\caption{Same as \reffig{plotmain} but with more restrictive assumptions on the 
running: (i) $-0.04<\as<-0.02$, (ii) $-0.02<\as<0$ and (iii) $0<\as<0.02$.}
\label{plotv1v2}
\end{figure}

The main result is displayed in \reffig{plotmain}, for which we assume 
$-0.04<\as<0.02$. For each order in the expansion of the potential there is a 
line which represents the boundary between the region where it is possible to 
build a working model (shaded) and the region where single-field inflation is 
excluded. The dots are independent samples from the Monte-Carlo Markov chain 
used in Ref.~\cite{LL03} to fit \textsc{wmap} and 2d\textsc{f}, and thus 
represent models providing a good fit to the perturbation data. We also plot the 
line $\e_2=0$ to compare with the naive expectation that only models with 
$\epsilon_2>0$ are suitable candidates for violating slow-roll (since by 
definition $\de\e_1/\de N=\e_1\e_2$).

In \reffig{plotv1v2} we make more restrictive assumptions on the running, with 
the three graphs showing the cases (i), (ii) and (iii) as described in 
Section~\ref{methodology}. Figure~\ref{plotnsm1r} shows the same information but 
displayed in the space of observable parameters, that is $n_\mathrm{S}-1$ and 
$R$. We comment on small-field and large-field inflation separately.

\subsection{Small-field inflation}

First, let us consider potentials with a negative second derivative $V''<0$, 
which are found in the lower part of each of the panels in \reffig{plotmain} and 
\reffig{plotv1v2}, and at the left lower corner of the panels in 
\reffig{plotnsm1r}. We can see in \reffig{plotmain} that these models are 
currently not constrained by the upper bound $\Nmax$. Indeed, it is always 
possible to build a working model even when considering only a third-order 
expansion. This is because as long as its slope is not bounded outside the 
region probed by observations, nothing prevents the potential from steepening 
enough in order to violate slow-roll before the number of $e$-foldings reaches 
$\Nmax$.

Even when tightening our assumptions on the running (see \reffig{plotv1v2}), 
models which fit the perturbation data remain almost unconstrained as long as 
the running is negative. However if the running is positive, panel~(iii), the 
third derivative prevents inflation from ending in time, and unless we use at 
least a fourth-order expansion, it is difficult to build a working model 
consistent with observations. Nevertheless, it is fair to say that so far 
small-field inflation is consistent with observations even when taking into 
account the constraint on $\Nmax$.

\subsection{Large-field inflation}

\begin{figure}[t!]
\includegraphics[scale=0.36,angle=-90]{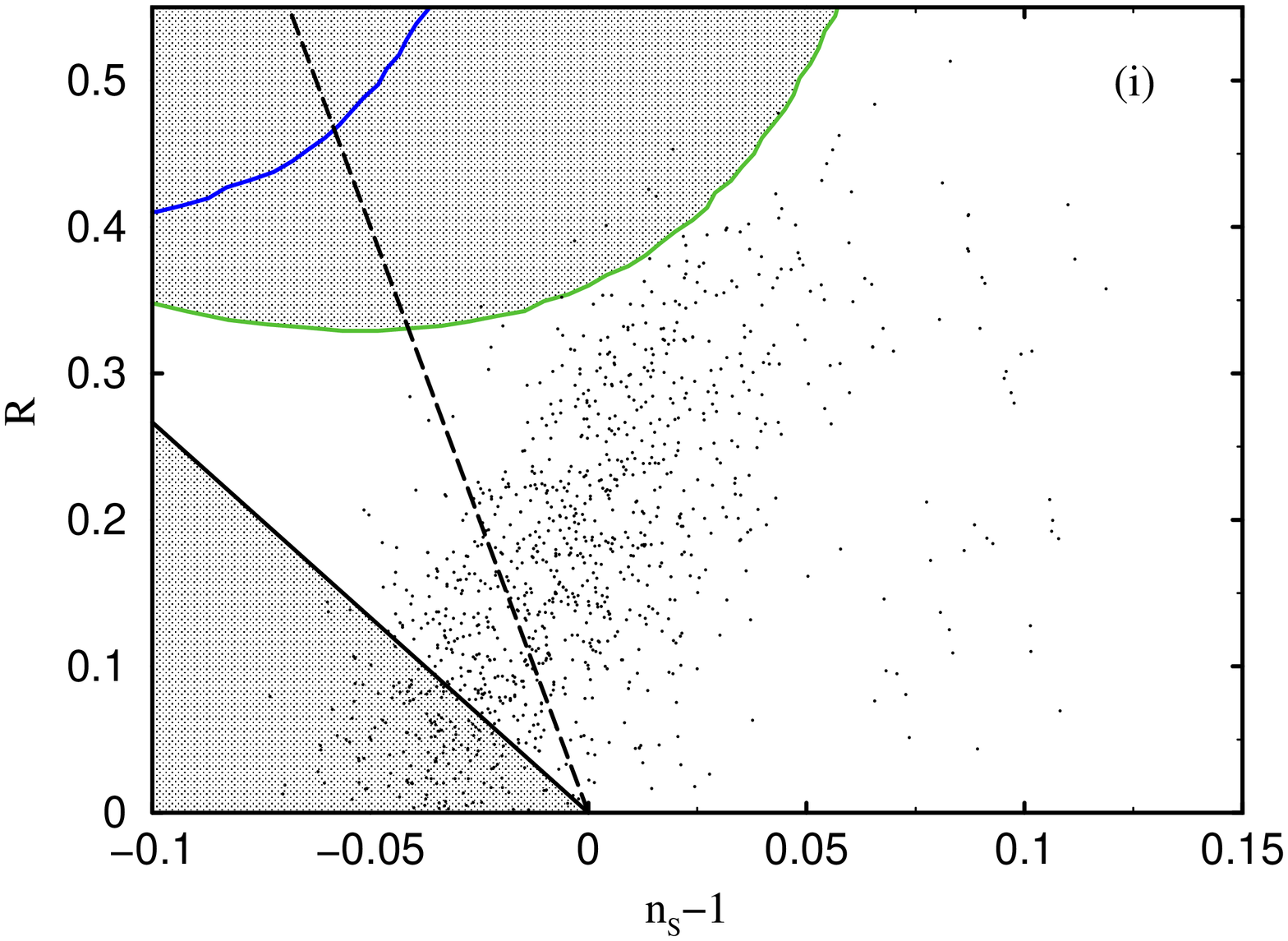}
\includegraphics[scale=0.36,angle=-90]{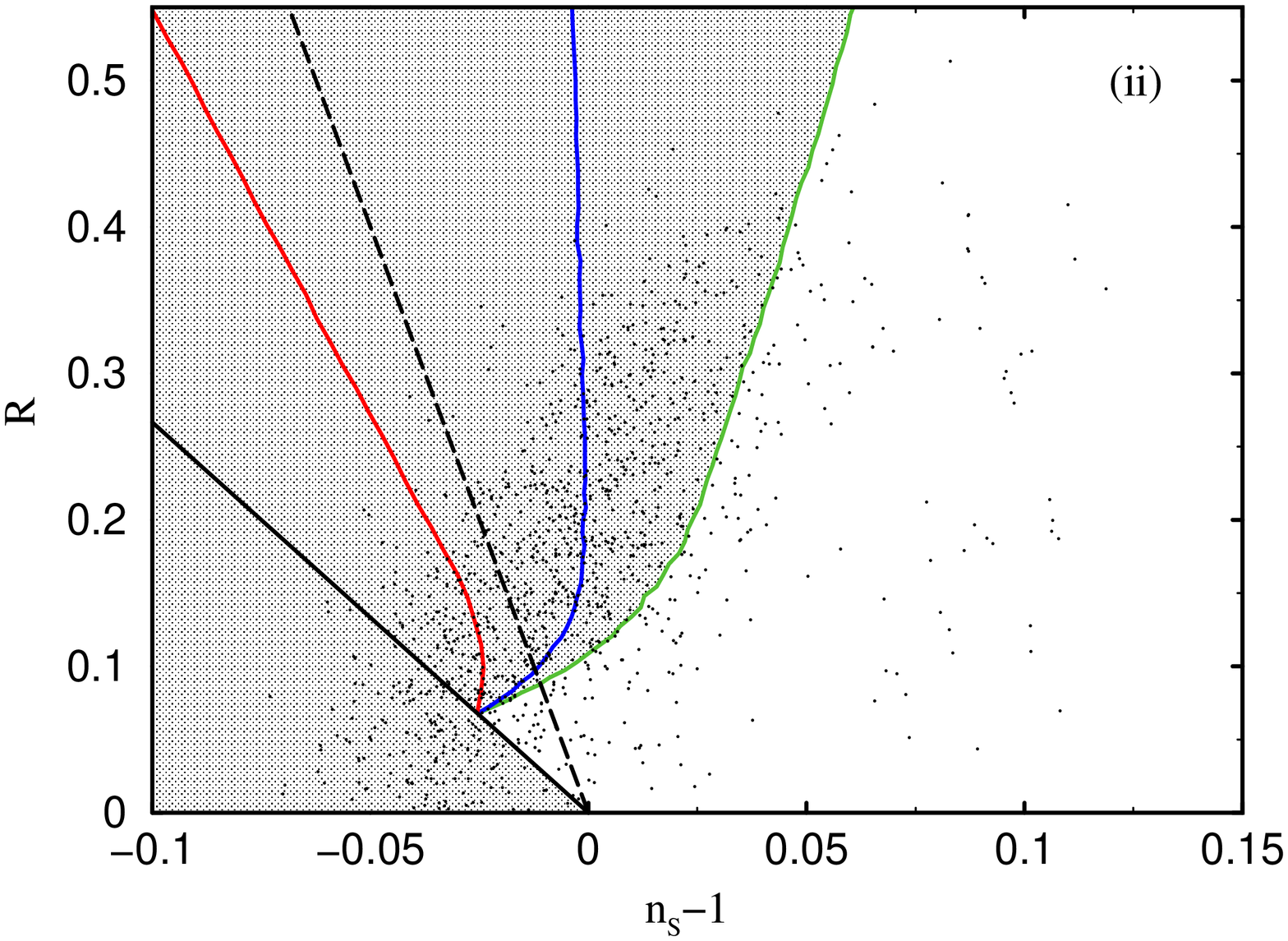}
\includegraphics[scale=0.36,angle=-90]{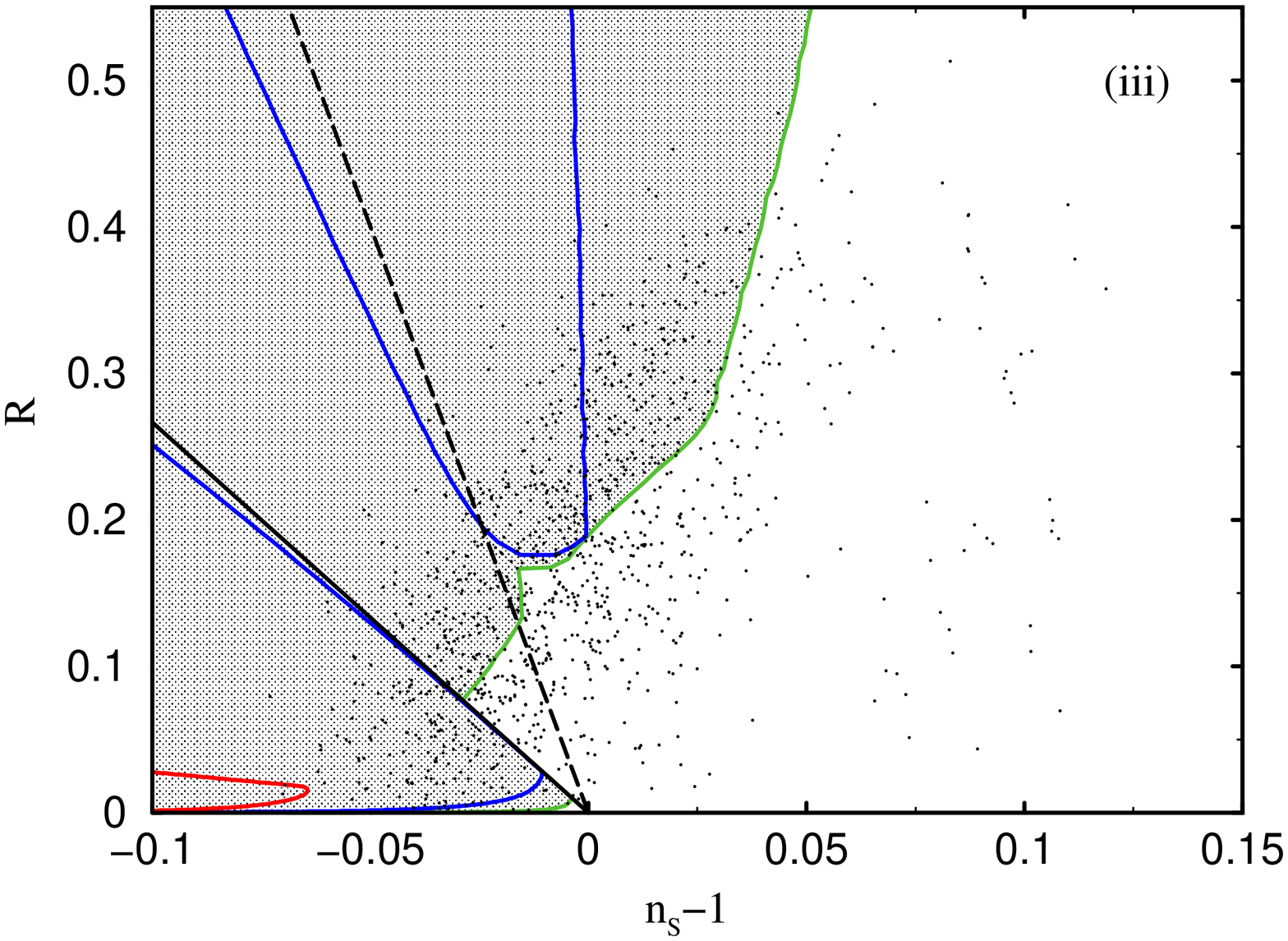}
\caption{Same as \reffig{plotv1v2} but in the $(n_\mathrm{S}-1)$--$R$ plane.}
\label{plotnsm1r}
\end{figure}

In the case of potentials with a positive curvature, the situation is very 
different. The modulus of the slope of the potential is bounded from above 
(because of the second rule) and the quickest, and therefore somewhat 
unrealistic, way to end inflation would be for the potential to become a linear 
potential as soon as the field leaves the observable region. We studied this 
kind of potential but it did not lead to any interesting constraint. Now, it is 
clear that the more derivatives we take into account, the more degrees of 
freedom we have available to shape the potential into a linear-like potential, 
and the extra degrees of freedom can conspire to build an extreme model in order 
to end inflation as quickly as possible. If such a model does not end inflation 
in time for given values of $V'/V$ and $V''/V$, then we can be sure that this 
pair of parameters is inconsistent with our class of single-field inflation 
models. Note that the allowed models lying close to the fifth-order boundary 
already require a certain amount of fine tuning between the different 
derivatives. One could of course expand the potential to sixth-order, but the 
resulting enlargement of the allowed region would be due to potentials that are 
even more linear-like and fine-tuned.

Figure~\ref{plotmain} shows that a significant fraction (around 30\%) of 
large-field models which fit the \textsc{wmap} and 2d\textsc{f} data sets are 
excluded by the need to end inflation in time. This proportion increases even 
more (to around 60\%) when considering only fourth-order expansion potentials 
and in the third-order case almost all of them (around 90\%) are excluded. It is 
somewhat unfortunate that the constraints coming from observations and from the 
need to end inflation lie in the same direction, but it is fair to say that 
large-field inflation models are under pressure.

This becomes clearer when we consider tighter constraints on the running. We can 
see in \reffig{plotv1v2} that a large negative running ($\as<-0.02$, panel (i)) 
is almost inconsistent with large-field inflation. This can be easily understood 
by looking at \refeq{v3}; rule~2 plays a major role by preventing the curvature 
of the potential from changing sign, and it will be difficult to build a working 
model unless $\e_1$ is large ($V'/V$ very negative). In the case of a positive 
running the possibility of a third-order working model is excluded and 
fourth-order models are difficult to achieve.

On the other hand, panel (ii) in \reffig{plotv1v2} ($-0.02<\as<0$) is exactly 
the same as \reffig{plotmain}. This means that it is much easier to end 
inflation if the running is between $-0.02$ and $0$ than if it is positive or 
more negative. In other words, for large-field inflation models, the running is 
tightly constrained by the need to end inflation. As a result, forthcoming 
surveys may rule out this class of inflation models.

From an observational point of view, when looking at \reffig{plotnsm1r} we see 
that our conditions clearly favor models with $n_\mathrm{S}<1$, and in 
particular we find that even $n_\mathrm{S}=1$ is hard to achieve unless $R$ is 
large.

\section{Discussion}
\label{discussion}

We have been motivated by the flow-equation formalism of Easther and 
Kinney~\cite{EK} to study the idea of randomly generating a large class of 
slow-roll inflation models in order to make a comparison with the increasingly 
restrictive observational constraints. However, as explained in Ref.~\cite{L03}, 
the flow equation formalism does not incorporate the underlying inflationary 
physics via the Euler-Lagrange equation. In our procedure this has been 
essential since we wanted to place a constraint on the qualitative shape of the 
inflation potential (via our rule 2).

Nevertheless, it is worth comparing with the results of Refs.~\cite{Petal,KKMR}
which used the flow-equation formalism. First of all, both of those papers have 
included the running as a parameter when generating their observational 
constraints. As a result, the observationally favored region in the 
$(n_\mathrm{S}-1)$--$R$ plane is enlarged, giving the effect that the 
flow-equation formalism currently picks out a small preferred region. Compared 
with observational constraints with no running, the flow-equation formalism 
actually generates a large class of models covering almost all of the observably 
favored region. Our method has generated a more restricted ensemble of inflation 
models, and from this perspective it can be considered a small step forward. 
Moreover, we have not tried to display any distribution of models, but instead 
just defined regions compatible with our class of single-field inflation models, 
arguing that the models near the edges of these regions are in some sense 
already fine tuned. This presentation has also allowed us to clarify the effect 
of adding further derivatives to our expansion of the potential.

Broadly speaking, we found it very easy to construct working models with 
$V''/V<0$, whereas for models with $V''/V>0$ the situation is more complex. 
Specifically, we showed that a lower limit on the amplitude of the slope of the 
potential does persist in the region classified as large-field inflation, 
analogous to the lower limit recently used to put pressure on the $\phi^\alpha$ 
inflation models \cite{LL03}. This means that the upper limit on $\Ninf$ does 
exert some pressure on inflation model building efforts. In addition, we showed 
that our constraints have a strong dependence on the running of the spectral 
index as it determines the value of the third derivative. From an observational 
point of view, we found that single-field inflation models can give 
$n_\mathrm{S}>1$, but only with a large value of $R$, which is expected to be 
constrained by upcoming observations.

To summarize, while small-field models are poorly constrained by the maximum 
number of $e$-foldings, we can see a certain tension against our large-field 
models and forthcoming observations may actually rule them out. 
Obviously some fundamental theory could be responsible for a potential with an 
unexpected shape, but for studying phenomenological models our assumptions seem 
reasonable. Finally we must stress that $\Nmax$ is an upper bound and knowing 
details about the reheating process may lower that bound and lead to even more 
constraining results.


\begin{acknowledgments}
M.M.~was supported by the Fondation Barbour, the Fondation Wilsdorf, the 
Janggen-P\"{o}hn-Stiftung and the \textsc{ors} Awards Scheme, S.M.L.~by the 
\textsc{eu cmbnet} network and A.R.L.~in part by the Leverhulme Trust.
\end{acknowledgments}


\end{document}